\documentclass[twocolumn,showpacs,pra,aps]{revtex4}

\usepackage{graphicx}
\usepackage{bm}
\usepackage{amsmath}
\usepackage{amssymb}
\usepackage{latexsym}
\usepackage{exscale}
\usepackage{natbib}

\newcommand{\ten}[1]{\mbox{\textbf{\textsf{#1}}}}

\newcommand{\mycomment}[1]{}
\newcommand{\Nabla}{\bm{\nabla}}
\newcommand{\Lnabla}{\overset{\leftarrow}{\Nabla}}

\newcommand{\D}{\mathrm{d}}

\newcommand{\dif}{\mathrm{d}}
\newcommand{\mi}{i}
\newcommand{\me}{e}
\newcommand{\sprod}{\cdot}
\newcommand{\tprod}{}
\newcommand{\vprod}{\times}
\newcommand{\trace}{{\operatorname{Tr}}}
\newcommand{\excess}{{\star}}
\newcommand{\grad}{\bm{\nabla}}
\newcommand{\curl}{\bm{\nabla}\times}
\newcommand{\dual}{{\circledast}}

\begin{document}

\title{Casimir--Polder interaction between an atom and a small
magnetodielectric sphere}

\author{Agnes Sambale}

\affiliation{Theoretisch--Physikalisches Institut,
Friedrich--Schiller--Universit\"at Jena,
Max--Wien-Platz 1, D-07743 Jena, Germany}

\author{Stefan Yoshi Buhmann}
\author{Stefan Scheel}
\affiliation{Quantum Optics and Laser Science, Blackett Laboratory,
Imperial College London, Prince Consort Road,
London SW7 2BW, United Kingdom}

\date{\today}

\begin{abstract}
On the basis of macroscopic quantum electrodynamics and
point-scattering techniques, we derive a closed expression for the
Casimir--Polder force between a ground-state atom and a small
magnetodielectric sphere in an arbitrary environment. In order to
allow for the presence of both bodies and media, local-field
corrections are taken into account. Our results are compared with the
known van der Waals force between two ground-state atoms. To
continuously interpolate between the two extreme cases of a single
atom and a macroscopic sphere, we also derive the force between an
atom and a sphere of variable radius that is embedded in an Onsager
local-field cavity. Numerical examples illustrate the theory.
\end{abstract}


\maketitle


\section{Introduction}
\label{Intro}

Van der Waals (vdW) dispersion forces are effective electromagnetic
forces that arise between polarisable objects as a consequence of
correlated quantum fluctuations
\cite{0608,0375,0620,0487,0323,0322,Review2006,0832,0853,0868}.
Postulated as early as 1873 by van der Waals in order to account for
deviations from the ideal gas law \cite{vdW}, they were theoretically
understood only much later in 1930, when London derived them from the
electrostatic Coulomb interaction of charge fluctuations \cite{0374}.
He found that two atoms of polarisabilities $\alpha_A(\omega)$ and
$\alpha_B(\omega)$ at a distance $r_{AB}$ are subject to a vdW
potential
\begin{equation}
\label{London}
U(r_{AB})=-\frac{3\hbar}{16\pi^3\varepsilon_0^2r_{AB}^6}
 \int_0^\infty\dif\xi\,\alpha_{A}(\mi\xi)\alpha_{B}(\mi\xi).
\end{equation}
Important progress was again made in 1948 by Casimir and Polder who
included fluctuations of the transverse electromagnetic field to
obtain a full quantum electrodynamic description of dispersion forces
\cite{0030}. Their result reduces to the London formula in the 
electrostatic limit, but is given by
\begin{equation}
\label{CasimirPolder}
U(r_{AB})
 =-\frac{23\hbar c\alpha_{A}(0)\alpha_{B}(0)}
 {64\pi^3\varepsilon_0^2 r_{AB}^7}
\end{equation}
for distances much larger than the relevant atomic transition
wavelengths.

Casimir arrived at his famous results while studying the properties
of colloidal solutions \cite{Lambrecht02}. This illustrates the
importance of dispersion forces to colloid science, which deals with
the (inter alia vdW type) interactions between small clusters of
particles in free space \cite{Gatica2005} and more often with the
different forces in colloidal suspensions. For example, (attractive)
dispersion forces between spherical micro- and macro objects embedded
in a liquid \cite{Kim2007} usually diminish the stability of such
suspensions and may even cause clustering or flocculation
\cite{Thomas1999}. The introduction of small amounts of highly charged
nanoparticles gives rise to competing repulsive forces, thus balancing
the stability of the suspension \cite{Liu2004}. Stable mechanical
suspensions might also be created with fluid-separated macro-objects
such as eccentric cylinders by means of repulsive dispersion forces
\cite{Rodriguez2008}. Note that in addition to dispersion and
electrostatic forces, critical Casimir forces due to concentration
fluctuations \cite{Hertlein08}, chemical effects such as hydration,
solvation and hydrophobic forces as well as steric repulsion
\cite{Gregory1993,Liang2007} and depletion \cite{Fuchs2002} also
influence the interaction between the colloidal particles.

Dispersion forces play a similar role in biology, where they
contribute to the organisation of molecules
\cite{0371,Israelachvili,0474}, cell adhesion
\cite{0371,0359,0495,Israelachvili} and the interaction of molecules
with cell membranes \cite{0359,Israelachvili}. They are further of
interest in atomic force microscopy \cite{Bruch2005}. 

\begin{figure}[!t!]
\begin{center}
\includegraphics[width=\linewidth]{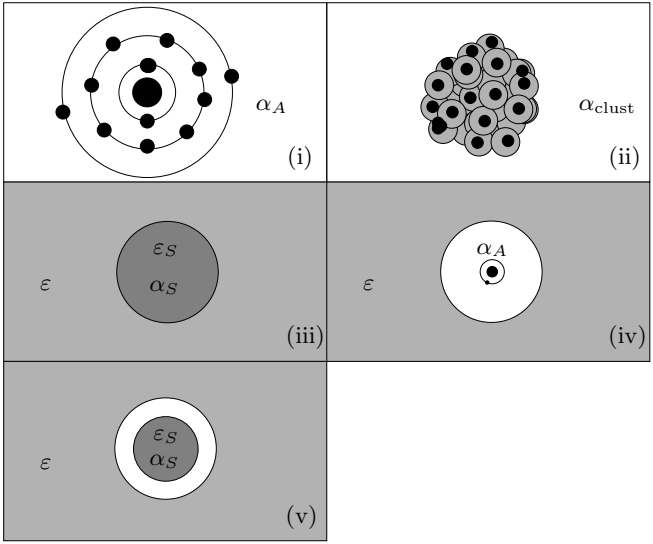}
\end{center}
\caption{
\label{Models}
Models of polarisable systems: (i) Neutral arrangement of point
charges (atom), (ii) Cluster of atoms, (iii) Dielectric sphere inside
medium, (iv), Atom in empty cavity surrounded by medium (Onsager real
cavity model), (v) Dielectric sphere in empty cavity surrounded by
medium.
}
\end{figure}

A large variety of models have been used in the past to study vdW
forces between small polarisable objects \cite{Kim2007}. Interacting
atoms have been studied on a microscopic level as neutral
arrangements of point charges \cite{0374,0030} as sketched in
Fig.~\ref{Models}(i). Larger systems can be treated by considering
collections of such polarisable point objects [Fig.~\ref{Models}(ii)],
where often a pairwise-sum method is employed \cite{0693}. 
Investigations of the polarisability of $N$-atom nanoclusters of
various sizes and shapes have shown that an additive relation
$\alpha_\mathrm{clust}= N\alpha_A$ does not hold in general, but is
valid for spherical clusters \cite{Kim2005}. Microscopic approaches
have to be contrasted with macroscopic descriptions where continuous
objects of polarisable matter are characterised by their permittivity
[Fig.~\ref{Models}(iii)], and an intervening medium can be accounted 
for in the same spirit. One commonly distinguishes the additive
Hamaker method \cite{Hamaker37} from the more elaborate Lifshitz
theory which includes many-body interactions
\cite{Lifshitz56,ReviewLifshitz}.

A hybrid approach consists in a
microscopic treatment of interacting atoms, combined with a
macroscopic description of an intervening medium. To reconcile the
microscopic and macroscopic pictures, local-field effects are included
by assuming the atoms to be surrounded by small free-space cavities,
an approach known as the Onsager real-cavity model \cite{0488}, see
Fig.~\ref{Models}(iv). As found by studying the behaviour of the
classical Green tensor $\ten{G}$ for the electromagnetic field in
conjunction with the real-cavity model \cite{0489}, the vdW potential
of two atoms at positions $\mathbf{r}_A$, $\mathbf{r}_B$ in an
arbitrary environment can be given as \cite{Sambale2007}
\begin{align}
\label{vdWloc}
U(\mathbf{r}_A,\mathbf{r}_B)
 =&-\frac{\hbar\mu_0^2}{2\pi}\int_0^\infty
 \dif\xi\,\xi^4\alpha_A(\mi\xi)\alpha_B(\mi\xi)\nonumber\\
&\times
 \biggl[\frac{3\varepsilon_A(\mi\xi)}{2\varepsilon_A(\mi\xi)+1}
 \biggr]^2
 \biggl[\frac{3\varepsilon_B(\mi\xi)}{2\varepsilon_B(\mi\xi)+1}
 \biggr]^2 \nonumber\\
&\times\trace\bigl[
 \ten{G}(\mathbf{r}_A,\mathbf{r}_B,\mi\xi)\sprod
 \ten{G}(\mathbf{r}_B,\mathbf{r}_A,\mi\xi)\bigr],
\end{align}
where local-field correction factors explicitly appear.

To compare the microscopic, local-field corrected approach of
Fig.~\ref{Models}(iv) with the macroscopic one shown in
Fig.~\ref{Models}(iii), we will in this work study the
Casimir--Polder (CP) interaction of a ground-state atom with a
magnetodielectric sphere of variable size in the presence of arbitrary
magnetoelectric background media on basis of macroscopic quantum
electrodynamics (QED). Our considerations are related to free-space
results obtained earlier for the interaction of an atom with curved
surfaces \cite{Messina2009}, dielectric \cite{0017} and perfectly
conducting spheres \cite{Taddei2009}; cf.~also the nonretarded vdW
potential of a ground-state atom inside and outside a dielectric or
metallic spherical shell as calculated in Ref.~\cite{Boustimi2000}. In
addition, we will consider the interaction of an atom with a
magnetodielectric sphere inside an Onsager cavity
[Fig.~\ref{Models}(v)]. By changing the radius of the sphere, this 
construction will allow us to study the transition between a
microscopic point-like object and a macroscopic one.

The paper is organised as follows. In Sec.~\ref{basic}, we recall
the basic equations concerning ground-state CP potentials. In
Sec.~\ref{Green}, it is shown how the Green tensors of a
magnetodielectric full sphere or a sphere within an Onsager cavity can
be written as functions of the Green tensors of the environment
without the sphere. The results are then used to study the atom-sphere
potentials, which are compared with the vdW interaction between two
ground-state atoms (Sec.~\ref{Pot}). As an example, we evaluate the 
interaction between an atom and molecules of different sizes in a bulk
medium. A summary is given in Sec.~\ref{Sum}.


\section{Atom--sphere interaction}
\label{basic}

In order to derive an expression for the CP interaction between a
(ground-state) atom and a magnetodielectric sphere in the presence of
an arbitrary medium environment, we start from the familiar formulas
for the electric ($U_e$) and magnetic ($U_m$) CP potentials of a
ground-state atom of polarisability 
\begin{equation}
\label{eq:alpha}
\alpha_A(\omega)
=\lim_{\epsilon\to 0}\frac{2}{3\hbar}\sum_k
\frac{\omega_{k0}|\mathbf{d}_{0k}|^2}
{\omega_{k0}^2-\omega^2-\mi\omega\epsilon}
\end{equation}
($\omega_{k0}$: transition frequencies; $\mathbf{d}_{0k}$: electric
dipole matrix elements) and magnetisability
\begin{equation}
\label{eq:beta}
\beta_A(\omega)
=\lim_{\epsilon\to 0}\frac{2}{3\hbar}\sum_k
\frac{\omega_{k0}|\mathbf{m}_{0k}|^2}
{\omega_{k0}^2-\omega^2-\mi\omega\epsilon}
 \end{equation}
($\mathbf{m}_{0k}$: magnetic dipole matrix elements) that is placed at
an arbitrary position $\mathbf{r}_A$ within an environment of locally
and linearly responding magnetoelectric bodies/media [characterised by
their permittivity $\varepsilon(\mathbf{r}_A,\omega)$ and permeability
$\mu(\mathbf{r}_A,\omega)$] \cite{Review2006,Sambale2007,Safari2008}:
\begin{multline}
\label{Ue}
U_e(\mathbf{r}_A)=\frac{\hbar\mu_0}{2\pi}\int _0 ^\infty \mathrm{d}
\xi\, \xi^2 \alpha_A(i\xi)
\Bigl[\frac{3\varepsilon_A(i\xi)}{2\varepsilon_A(i\xi)+1}\Bigr]^2\\
\times\mathrm{Tr}\ten{G}^{(1)}(\mathbf{r}_A,\mathbf{r}_A,i\xi),
\end{multline}
\begin{multline}
\label{Um}
U_m(\mathbf{r}_A)=\frac{\hbar\mu_0}{2\pi}\int _0 ^\infty
 \mathrm{d}\xi\, \beta_A(i\xi)
\Bigl[\frac{3}{2\mu_A(i\xi)+1}\Bigr]^2\\
\times\mathrm{Tr}\bigl[\bm{\nabla}
\times\ten{G}^{(1)}(\mathbf{r}_A,\mathbf{r}_A,i\xi)\times
\overleftarrow{\bm{\nabla}}'
\bigr],
\end{multline}
[$\varepsilon_A(\omega)\equiv \varepsilon(\mathbf{r}_A,\omega)$,
$\mu_A(\omega)\equiv \mu(\mathbf{r}_A,\omega)$]. These expressions
explicitly allow for the atom to be embedded in a medium environment
where the relevant local-field corrections have been accounted for via
the Onsager real-cavity model \cite{0488,0489}. The scattering Green
tensor $\ten{G}^{(1)}(\mathbf{r},\mathbf{r}',\omega)$ fully accounts
for the position, size, and shape of all bodies and media as well as
their magnetoelectric properties and is defined by the differential
equation
\begin{equation}
\label{nabla}
\Bigl[\Nabla \times \frac{1}{\mu(\mathbf{r},\omega)}\Nabla
\times-\frac{\omega^2}{c^2}\varepsilon(\mathbf{r},\omega)\Bigr]\ten{G}
(\mathbf{r},\mathbf{r}',\omega)=\bm{\delta}(\mathbf{r}-\mathbf{r}')
\end{equation}
with the condition
$\ten{G}(\mathbf{r},\mathbf{r}',\omega)\rightarrow\ten{0}$ for 
$|\mathbf{r}-\mathbf{r}'|\rightarrow\infty$. In this work, the body
interacting with the atom is a magnetodielectric sphere whose Green
tensor we will analyse in the following.


\subsection{Decomposition of the Green tensor}
\label{Green}

Two methods may be envisaged to study the CP potential of an atom in
the presence of a magnetodielectric sphere and an arbitrary
environment of additional bodies and media. First, one could work with
the Green tensor of the combined sphere--environment system
[Fig.~\ref{Decomp}(i)] directly which may be very complicated; even if
an analytical expression of the Green tensor is known, the resulting
expressions for the potentials are hard to evaluate and will not allow
for an explicit discussion of the influence of the sphere. In this
work, we therefore follow a second, alternative approach: We show
how the Green tensor of the full arrangement including the sphere,
$\ten{G}^{(1)}_S(\mathbf{r},\mathbf{r},\omega)$, can be related to the
Green tensor without the sphere [Fig.~\ref{Decomp}(ii)],
$\ten{G}^{(1)}(\mathbf{r},\mathbf{r},\omega)$ describing only the
background environment. To establish such a relation, we use methods
similar to those developed for studying local-field corrections
\cite{0489,Sambale2007,Safari2008}. The crucial assumption
for using these point-scattering techniques is that the
effective radius of the sphere is small compared with the relevant
wavelengths of the electromagnetic field.


\subsubsection{Full sphere}
\label{sub2l}

Assuming that the functions
$\varepsilon(\mathbf{r},\omega)$ and $\mu(\mathbf{r},\omega)$
describe the magnetoelectric properties of the environment, the
introduction of a homogeneous magnetodielectric sphere with radius $R$
centred at $\mathbf{r}_S$, and with permittivity
$\varepsilon_S(\omega)$ and permeability $\mu_S(\omega)$ will lead to
the new functions 
\begin{equation}
\label{eps}
\varepsilon_S(\mathbf{r},\omega), \mu_S(\mathbf{r},\omega)=
\begin{cases}
\varepsilon_S(\omega),\mu_S(\omega) & \mbox{for }
|\mathbf{r}-\mathbf{r}_S|\leq R,\\
\varepsilon (\mathbf{r},\omega),\mu (\mathbf{r},\omega) &
\mbox{elsewhere}.
\end{cases}
\end{equation}
The Green tensor $\ten{G}^{(1)}_S(\mathbf{r},\mathbf{r},\omega)$ of
sphere plus environment is hence the solution to the differential
equation~(\ref{nabla}) with $\varepsilon_S(\mathbf{r},\omega)$
and $\mu_S(\mathbf{r},\omega)$ in place of  
$\varepsilon(\mathbf{r},\omega)$ and $\mu(\mathbf{r},\omega)$.

\begin{figure}[!t!]
\begin{center}
\includegraphics[width=\linewidth]{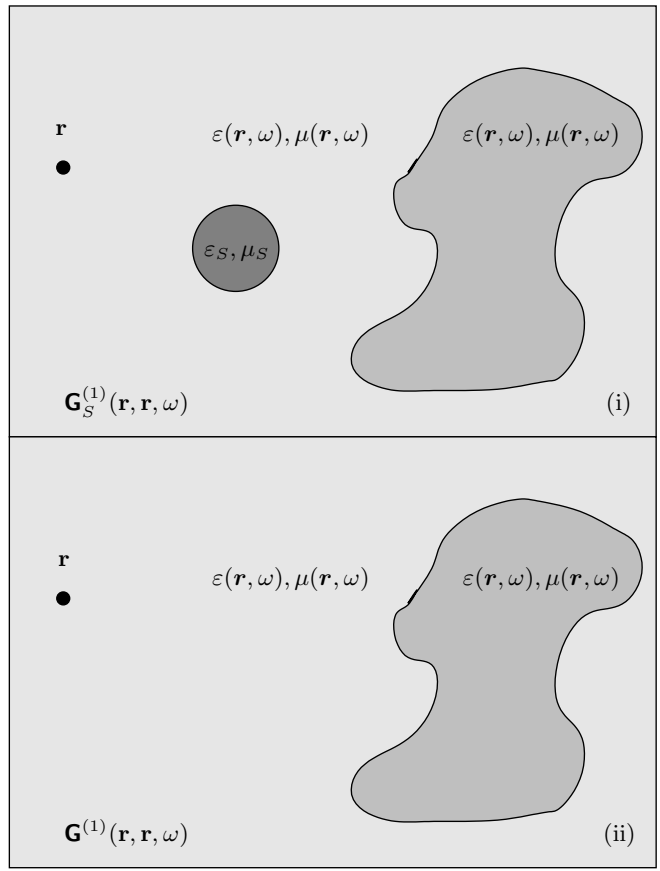}
\end{center}
\caption{
\label{Decomp}
(i) Green tensor of the combined system of magnetodielectric sphere
and arbitrary background environment, (ii) Green tensor of the same
system without the sphere. 
}
\end{figure}
As discussed in Refs.~\cite{0489,Sambale2007}, for a small sphere it
is sufficient to consider first the special case of a homogeneous,
bulk environment, which will then be generalised to arbitrary
environments at the end of the section. The required scattering Green
tensor of the sphere with centre $\mathbf{r}_S=\mathbf{0}$ inside a
bulk medium of permittivity $\varepsilon(\omega)$ and permeability
$\mu(\omega)$ can be written in the form \cite{Li1994}
\begin{multline}
\label{G}
\ten{G}_S^{(1)}(\mathbf{r},\mathbf{r}',\omega)
=
\frac{i\mu k}{4\pi}
\sum _{p=\pm}\sum _{l=1}^\infty \sum _{m=0}^n
(2-\delta_{m0})\\
\times\frac{2l+1}{l(l+1)}\frac{(l-m)!}{(l+m)!}
\bigl[B_l^M
\mathbf{M}_{lmp}(k,\mathbf{r})\tprod\mathbf{M}_{lmp}(k,\mathbf{r}')\\+
B_l^N \mathbf{N}_{lmp}(k,\mathbf{r})
\tprod\mathbf{N}_{lmp}(k,\mathbf{r}')\Bigr]
\end{multline}
($k=\sqrt{\varepsilon\mu}\,\omega/c$)
where $\mathbf{M}_{lmp},\mathbf{N}_{lmp}$ denote even ($+$) and odd
($-$) spherical vector wave functions with total angular momentum $l$
and $z$-projection $m$; $B_l^{M,N}$ are the associated coefficients
for reflection at the surface of the sphere. Explicit forms for
$\mathbf{M}_{lmp}$, $\mathbf{N}_{lmp}$, and $B_l^{M,N}$ can be found
in Refs.~\cite{Li1994,Chew1995}.

In the limit of a small sphere with $|k_SR|,|kR|\ll 1$
($k_S=\sqrt{\varepsilon_S\mu_S}\,\omega/c$), we have
\begin{equation} 
\label{BMN}
B_l^{M,N}={\cal O}\biggl(\frac{\omega R}{c}\biggr)^{2l+1},
\end{equation}
so the dominant contribution to the Green tensor is due to the
$l=1$ terms. The respective vector wave functions are given by
\begin{multline}
\label{M}
\mathbf{M}_{1m\pm}(k,\mathbf{r})=\mp\frac{m}{\sin\theta}\,
 h_1^{(1)}(kr)P_1^m(\cos \theta)
\begin{array}{l}
	\sin m\phi\\
	\cos m \phi 
\end{array}
\mathbf{e}_\theta\\
-h_1^{(1)}(kr)\,\frac{\D P_1^m(\cos \theta)}{\D \theta}\,
\begin{array}{l}
	\cos m\phi\\
	\sin m \phi 
\end{array}
\mathbf{e}_\phi,
\end{multline}
\begin{multline}
\label{N}
\mathbf{N}_{1m\pm}(\mathbf{r})=2\,\frac{h_1^{(1)}(kr)}{kr}\,
 P_1^m(\cos\theta)
\begin{array}{l}
	\cos m\phi\\
	\sin m \phi 
\end{array}\mathbf{e}_r\\
+\frac{1}{kr}\,\frac{\D [kr h_1^{(1)}(kr)]}{d(kr)}\,
 \frac{\D P_1^m(\cos 
\theta)}{\D \theta}\,
\begin{array}{l}
	\cos m\phi\\
	\sin m \phi 
\end{array}\mathbf{e}_\theta\\
\mp \frac{m}{\sin \theta}\, P_1^m(\cos \theta)\,\frac{1}{kr}\,\frac{\D
[kr h_1^{(1)}(kr)]}{d(kr)}\,
\begin{array}{l}
	\sin m\phi\\
	\cos m \phi 
\end{array}\mathbf{e}_\phi
\end{multline}
where the $P_1^m(x)$ are associated Legendre polynomials and
$h_1^{(1)}(x)$ is a spherical Hankel function of the first kind.
The $l=1$ reflection coefficients in the small-sphere limit
$|k_SR|,|kR|\ll 1$ are given by
\begin{gather}
\label{BM}
B_{1}^M=\frac{2\mi}{3}
 \biggl(\sqrt{\varepsilon\mu}\,\frac{\omega R}{c}\biggr)^3 
 \frac{\mu_S-\mu }{\mu_S+2\mu}\,,\\
\label{BN}
B_{1}^N=\frac{2\mi}{3}
 \biggl(\sqrt{\varepsilon\mu}\,\frac{\omega R}{c}\biggr)^3
 \frac{\varepsilon_S-\varepsilon}
 {\varepsilon_S+2\varepsilon}\,.
\end{gather}
We can further evaluate the $p$ and $m$ sums for $l=1$ using
$P_1^0(\cos\theta)=\cos\theta$ and $P_1^1(\cos\theta)=-\sin\theta$
to obtain
\begin{multline}
\label{MM}
\sum _{p=\pm 1}
 \sum _{m=0}^1
(2-\delta_{m0})
\frac{(1-m)!}{(1+m)!}
\mathbf{M}_{1mp}(\mathbf{r})\tprod\mathbf{M}_{1mp}(\mathbf{r})\\
=h^2[\ten{I}-\mathbf{e}_r\tprod\mathbf{e}_r]
\end{multline}
and
\begin{multline}
\label{NN}
\sum _{p=\pm 1}
 \sum _{m=0}^1
(2-\delta_{m0})
\frac{(1-m)!}{(1+m)!}
\mathbf{N}_{1mp}(\mathbf{r})\tprod\mathbf{N}_{1mp}(\mathbf{r})\\
 =\frac{h'^2}{(kr)^2}\,\ten{I}
 +\frac{4h^2-h'^2}{(kr)^2}\,\mathbf{e}_r\tprod\mathbf{e}_r,
\end{multline}
with the notation
$h\equiv h_1^{(1)}(kr)=-\mi[1-\mi kr]\me^{\mi kr}/(kr)^2$ and
$h'\equiv\dif[krh_1^{(1)}(kr)]/\dif(kr)$. Substituting
these expressions into Eq.~(\ref{G}), the (equal-position)
scattering Green tensor of a small sphere becomes
\begin{multline}
\label{Gsresult}
\ten{G}^{(1)}_S(\mathbf{r},\mathbf{r},\omega)\\
 =\frac{\mu\me^{2\mi kr}}{4\pi k^2r^6}
 \biggl\{\left[1\!-\!2\mi kr\!-\!3(kr)^2\!+\!2\mi(kr)^3\!
 +\!(kr)^4\right]\ten{I}\\
 +\left[3\!-\!6\mi kr\!-\!(kr)^2\!-\!2\mi(kr)^3\!-\!(kr)^4\right]
 \mathbf{e}_r\tprod\mathbf{e}_r\biggr\}\\
\times\frac{\varepsilon_S-\varepsilon}{\varepsilon_S+2\varepsilon}\,
 R^3
 +\frac{\mu\me^{2\mi kr}}{4\pi r^4}
 \left[1\!-\!2\mi kr\!-\!(kr)^2\right]\\
\times(\ten{I}-\mathbf{e}_r\tprod\mathbf{e}_r)
 \frac{\mu_S-\mu}{\mu_S+2\mu}\,R^3.
\end{multline}

Next, we relate our result to the Green tensor of the bulk
medium without the sphere (cf., e.g. Ref.~\cite{Chew1995}),
\begin{multline}
\label{bulk}
\ten{G}(\mathbf{r},\mathbf{r}',\omega)
=-\frac{\mu\me^{\mi k\rho}}{4\pi k^2\rho^3}
\Bigl\{[1-\mi k\rho-(k\rho)^2)]\ten{I}\\
-[3-3ik\rho-(k\rho)^2]\mathbf{e}_\rho\tprod\mathbf{e}_\rho\Bigr\}
\end{multline}
which is valid for $\mathbf{r}\neq\mathbf{r}'$, where
$\bm{\rho}=\mathbf{r}-\mathbf{r}'$, $\rho=|\bm{\rho}|$,
$\mathbf{e}_\rho=\bm{\rho}/\rho$. The Green tensor
$\ten{G}^{(1)}_S(\mathbf{r},\mathbf{r},\omega)$ of the small
sphere describes the propagation of the electric field from a source
at $\mathbf{r}$ to the sphere, its scattering from the sphere (a
polarisable and magnetisable point scatterer) at $\mathbf{r}_S=\bm{0}$
and its return to $\mathbf{r}$. It is therefore natural to try and
compose $\ten{G}^{(1)}_S$ from products of $\ten{G}$, which describes
the propagation of the electric field through the bulk medium to an
electric scatterer, and $\ten{G}\vprod\overleftarrow{\bm{\nabla}}$,
which describes its propagation to a magnetic scatterer. Indeed, from
Eq.~(\ref{bulk}) we find that
\begin{multline}
\label{bulkbulk}
\ten{G}(\mathbf{r},\mathbf{0},\omega)\cdot
\ten{G}(\mathbf{0},\mathbf{r},\omega)\\
 =\frac{\mu^2\me^{2\mi kr}}{16\pi^2 k^4r^6}
 \Bigl\{\left[1\!-\!2\mi kr\!-\!3(kr)^2\!+\!2\mi(kr)^3\!+\!(kr)^4\right]
 \ten{I}\\
 +\left[3\!-\!6\mi kr\!-\!(kr)^2\!-\!2\mi(kr)^3\!-\!(kr)^4\right]
 \mathbf{e}_r\tprod \mathbf{e}_r\Bigr\}
\end{multline}
and
\begin{multline}
\label{NabGbulk}
\ten{G}(\mathbf{r},\mathbf{r}_S,\omega)\times \Lnabla_S\sprod
\Nabla_S\times\ten{G}(\mathbf{r}_S,\mathbf{r},\omega)
|_{\mathbf{r}_S=\bm{0}}\\
=-\frac{\mu^2\me^{2\mi kr}}{16\pi^2r^4}
 \left[1\!-\!2ikr\!-\!(kr)^2\right]
 (\ten{I}-\mathbf{e}_r\tprod\mathbf{e}_r).
\end{multline}
A comparison with Eq.~(\ref{Gsresult}) shows that 
\begin{multline}
\label{Gsresultinbulk}
\ten{G}^{(1)}_S(\mathbf{r},\mathbf{r},\omega)
 =4\pi\varepsilon R^3\,
 \frac{\varepsilon_S-\varepsilon}{\varepsilon_S+2\varepsilon}\\
\times\frac{\omega^2}{c^2}\,\ten{G}(\mathbf{r},\mathbf{0},\omega)\cdot
 \ten{G}(\mathbf{0},\mathbf{r},\omega)
 -\frac{4\pi R^3}{\mu}\,\frac{\mu_S-\mu}{\mu_S+2\mu}\\
\times\ten{G}(\mathbf{r},\mathbf{r}_S,\omega)\vprod\Lnabla_S\sprod
 \Nabla_S\vprod
 \ten{G}(\mathbf{r}_S,\mathbf{r},\omega)|_{\mathbf{r}_S=\mathbf{0}}.
\end{multline}

Let us next consider a general background environment, which can
involve different media or bodies, as sketched in Fig.~\ref{Decomp}.
With the permittivity $\varepsilon(\mathbf{r},\omega)$ and permeability
$\mu(\mathbf{r},\omega)$ of the environment now being functions of
position, it is useful to introduce a notation for their values at the
position of the sphere,
$\varepsilon_\odot(\omega)\equiv\varepsilon(\mathbf{r}_S,\omega)$,
$\mu_\odot(\omega)\equiv\mu(\mathbf{r}_S,\omega)$. In addition to the
small-sphere limit $|k_SR|\ll 1$, we will assume the effective sphere
radius $\sqrt{\varepsilon_S\mu_S}R$ to be much smaller than the
distance from the sphere to any of the environment bodies. As
demonstrated in Refs.~\cite{0489,Sambale2007}, multiple scattering
between sphere and environment can then be safely neglected to within
leading order of $k_SR$ and a result of the
type~(\ref{Gsresultinbulk}) can be generalised from the bulk case
to an arbitrary environment by adding the scattering Green tensor
and replacing $\varepsilon\mapsto\varepsilon_\odot$ and
$\mu\mapsto\mu_\odot$. This can be formally proven by treating both
the sphere and the environment bodies via a Born expansion of the
Green tensor \cite{Buhmann2006b} and discarding those terms in the
Born series that involve multiple scattering between atom and
environment. We obtain
\begin{multline}
\label{Gsresultgeneral}
\ten{G}^{(1)}_S(\mathbf{r},\mathbf{r},\omega)
 =\ten{G}^{(1)}(\mathbf{r},\mathbf{r},\omega)\\
 +\frac{\varepsilon_\odot}{\varepsilon_0}\,\alpha_S^\excess\,
 \frac{\omega^2}{c^2}\,\ten{G}(\mathbf{r},\mathbf{r}_S,\omega)\cdot
 \ten{G}(\mathbf{r}_S,\mathbf{r},\omega)\\
 -\frac{\mu_0}{\mu_\odot}\,\beta_S^\excess
 \ten{G}(\mathbf{r},\mathbf{r}_S,\omega)\vprod\Lnabla_S\sprod
 \Nabla_S\vprod\ten{G}(\mathbf{r}_S,\mathbf{r},\omega)
\end{multline}
where we have introduced the polarisability
\begin{equation}
\label{alpha}
\alpha_S^\excess=4\pi\varepsilon_0R^3\,
 \frac{\varepsilon_S-\varepsilon_\odot}
 {\varepsilon_S+2\varepsilon_\odot}
\end{equation}
and the magnetisability 
\begin{equation}
\label{beta}
\beta_S^\excess=\frac{4\pi R^3}{\mu_0}\,
 \frac{\mu_S-\mu_\odot}{\mu_S+2\mu_\odot}
\end{equation}
of the sphere \cite{Xu1992}. Note that $\alpha_S^\excess$ is an excess
or effective polarisability \cite{Landau63,McLach1965} and describes
the electric response of the sphere with respect to that of the
surrounding medium. It can take positive or negative values, depending
on whether the sphere's permittivity is larger or smaller than that of
the medium. 

The relation~(\ref{Gsresultgeneral}) for $\ten{G}^{(1)}_S$ can
be used to calculate the electric CP potential~(\ref{Ue}). In order
to find the magnetic CP potential, Eq.~(\ref{Um}), we also require the
analogous relation for the magnetic Green tensor 
$\bm{\nabla}\vprod\ten{G}^{(1)}_S\vprod\overleftarrow{\bm{\nabla}}'$,
which can be obtained by duality arguments. An electric/magnetic
duality transformation $[\,\cdot\,]^\dual$ corresponds to a global
exchange of electric and magnetic properties, $\varepsilon^\dual=\mu$,
$\mu^\dual=\varepsilon$. As shown in
Refs.~\cite{Buhmann2008a,Buhmann2009}, this results in the
following changes of the Green tensor:
\begin{gather}
\label{dgftrans1}
\frac{\omega^2}{c^2}\,\ten{G}^\dual(\mathbf{r},\mathbf{r}',\omega)
 =-
 \frac{\curl\ten{G}(\mathbf{r},\mathbf{r}',\omega)\vprod
 \overleftarrow{\grad}'}
 {\mu(\mathbf{r},\omega)\mu(\mathbf{r}',\omega)},\\
\label{dgftrans2}
\curl\ten{G}^\dual(\mathbf{r},\mathbf{r}',\omega)\vprod
 \overleftarrow{\grad}'
=-\varepsilon(\mathbf{r},\omega)\,\frac{\omega^2}{c^2}\,
 \ten{G}(\mathbf{r},\mathbf{r}',\omega)
 \varepsilon(\mathbf{r}',\omega),\\
\label{dgftrans3}
\curl\ten{G}^\dual(\mathbf{r},\mathbf{r}',\omega)
=-\varepsilon(\mathbf{r},\omega)\,
 \frac{\ten{G}(\mathbf{r},\mathbf{r}',\omega)\vprod\overleftarrow{\grad}'}
 {\mu(\mathbf{r}',\omega)}\,,\\
\label{dgftrans4}
\ten{G}^\dual(\mathbf{r},\mathbf{r}',\omega)\vprod\overleftarrow{\grad}'
=-\frac{\curl\ten{G}(\mathbf{r},\mathbf{r}',\omega)}
 {\mu(\mathbf{r},\omega)}\,
 \varepsilon(\mathbf{r}',\omega)
\end{gather}
for $\mathbf{r}\neq\mathbf{r}'$. In addition, Eqs.~(\ref{alpha}) and
(\ref{beta}) imply that $\alpha_S^{\star\dual}=\beta_S^\star/c^2$ and
$\beta_S^{\star\dual}=c^2\alpha_S^\star$. Applying the duality
transformation to both sides of Eq.~(\ref{Gsresultgeneral}), one
obtains the required relation
\begin{multline}
\label{Gsmresultgeneral}
\nabla\!\vprod\!\ten{G}_S^{(1)}(\mathbf{r},\mathbf{r}',\omega)\!\vprod\!
 \overleftarrow{\grad}'|_{\mathbf{r}'=\mathbf{r}}\\
=\nabla\!\vprod\!\ten{G}^{(1)}(\mathbf{r},\mathbf{r}',\omega)\!\vprod\!
 \overleftarrow{\grad}'|_{\mathbf{r}'=\mathbf{r}}
 -\frac{\mu_0}{\mu_\odot}\,\beta_S^\excess\\
\times\grad\!\vprod\!
 \ten{G}(\mathbf{r},\mathbf{r}_S,\omega)\!\vprod\!\Lnabla_S
 \!\cdot\!\Nabla_S\!\vprod\!
 \ten{G}(\mathbf{r}_S,\mathbf{r}',\omega)\!\vprod\!
 \overleftarrow{\grad}'|_{\mathbf{r}'=\mathbf{r}}\\
 +\frac{\varepsilon_\odot}{\varepsilon_0}\,\alpha_S^\excess\,
 \frac{\omega^2}{c^2} 
 \grad\!\vprod\!\ten{G}(\mathbf{r},\mathbf{r}_S,\omega)\sprod
 \ten{G}(\mathbf{r}_S,\mathbf{r}',\omega)\!\vprod\!
 \overleftarrow{\grad}'|_{\mathbf{r}'=\mathbf{r}}\;.
\end{multline}
%


\subsubsection{Sphere inside an Onsager cavity}
\label{sub3L}

Next, we consider a homogeneous magnetodielectric sphere with radius
$R$ centred around $\mathbf{r}_S$, with permittivity
$\varepsilon_S(\omega)$ and permeability $\mu_S(\omega)$, which is not
in immediate contact with the surrounding medium, but placed inside a
small spherical cavity of radius $R_C$, also centred around
$\mathbf{r}_S$ [Fig.~\ref{Models}(v)]. This will enable us to compare
and interpolate between the homogeneous sphere placed inside a medium
(as considered in the previous section) and a local-field corrected
atom (i.e., a point-like polarisable system surrounded by a
cavity). The scattering Green tensor $\ten{G}^{(1)}_{S+C}$ of the
sphere plus cavity system in a homogeneous bulk medium is again given
by an equation of the form~(\ref{G}) where the reflection coefficients
now take a more complex form. In particular, in the limit of a small
effective cavity and sphere sizes $|k_SR|,|kR_C|\ll 1$, one has
\cite{Chew1995} 
\begin{align}
\label{BM3layer}
&B_1^M=\frac{2\mi}{3}\biggl(\sqrt{\varepsilon\mu}
 \,\frac{\omega}{c}\biggr)^3
 \biggl[R_C^3\,\frac{1\!-\!\mu}{1\!+\!2\mu}\nonumber\\
&\quad+\frac{9\mu R^3(\mu_S\!-\!1)/(2\mu\!+\!1)}
 {(\mu_S\!+\!2)(2\mu\!+\!1)+2(\mu_S\!-\!1)(1\!-\!\mu)
 R^3/R_C^3}\biggr],\\
\label{BN3layer}
&B_1^N=\frac{2i}{3}\biggl(\sqrt{\varepsilon\mu}\,
 \frac{\omega}{c}\biggr)^3
 \biggl[R_C^3\,\frac{1\!-\!\varepsilon}{1\!+\!2\varepsilon}
 \nonumber\\
&\quad+\frac{9\varepsilon R^3
 (\varepsilon_S\!-\!1)/(2\varepsilon\!+\!1)}
 {(\varepsilon_S\!+\!2)(2\varepsilon\!+\!1)
 +2(\varepsilon_S\!-\!1)(1\!-\!\varepsilon)
 R^3/R_C^3}\biggr]\,.
\end{align}

We can then follow exactly the same steps as in the previous
Sec.~\ref{sub2l}. We again arrive at Eqs.~(\ref{Gsresultgeneral}) and
(\ref{Gsmresultgeneral}) with $S+C$ in place of $S$. A comparison of
Eqs.~(\ref{BM}) and (\ref{BN}) with Eqs.~(\ref{BM3layer}) and
(\ref{BN3layer}) shows that the relevant excess polarisability and
magnetisability of the sphere plus cavity system are given by
\begin{multline}
\label{alphaSC}
\alpha_{S+C}^\excess=4\pi\varepsilon_0\biggl[R_C^3\,
 \frac{1\!-\!\varepsilon_\odot}{1\!+\!2\varepsilon_\odot}\\
+\frac{9\varepsilon_\odot R^3
 (\varepsilon_S\!-\!1)/(2\varepsilon_\odot\!+\!1)}
 {(\varepsilon_S\!+\!2)(2\varepsilon_\odot\!+\!1)
 +2(\varepsilon_S\!-\!1)(1\!-\!\varepsilon_\odot) 
 R^3/R_C^3}\biggr]
\end{multline}
and
\begin{multline}
\label{betaSC}
\beta_{S+C}^\excess=\frac{4\pi}{\mu_0}\,\biggl[
 R_C^3\,\frac{1\!-\!\mu_\odot}{1\!+\!2\mu_\odot}\\
+\frac{9\mu_\odot R^3(\mu_S\!-\!1)/(2\mu_\odot\!+\!1)}
 {(\mu_S\!+\!2)(2\mu_\odot\!+\!1)+2(\mu_S\!-\!1)(1\!-\!\mu_\odot)
 R^3/R_C^3}\biggr]\,.
\end{multline}
One can easily verify that, for $R=R_C$, Eqs.~(\ref{alphaSC}) and
(\ref{betaSC}) reduce to the results~(\ref{alpha}) and (\ref{beta})
for the full sphere, as expected.

By introducing the free-space polarisability and magnetisability of
the sphere
\begin{equation}
\label{alphafree}
\alpha_S=4\pi\varepsilon_0R^3\,
 \frac{\varepsilon_S-1}{\varepsilon_S+2}
\end{equation}
and 
\begin{equation}
\label{betafree}
\beta_S=\frac{4\pi R^3}{\mu_0}\,\frac{\mu_S-1}{\mu_S+2}
\end{equation}
as well as the excess polarisability and magnetisability of the cavity
\begin{equation}
\label{alphaC}
\alpha_C^\excess=4\pi\varepsilon_0R_C^3\,
 \frac{1-\varepsilon_\odot}{1+2\varepsilon_\odot}
\end{equation}
and 
\begin{equation}
\label{betaC}
\beta_C^\excess=\frac{4\pi R_C^3}{\mu_0}\,
 \frac{1-\mu_\odot}{1+2\mu_\odot}\;,
\end{equation}
we can write Eqs.~(\ref{alphaSC}) and (\ref{betaSC}) more
transparently as
\begin{gather}
\label{alphaSC2}
\alpha_{S+C}^\excess=\alpha_C^\excess
+\frac{\alpha_S}{\varepsilon_\odot}\,
 \biggl(\frac{3\varepsilon_\odot}{2\varepsilon_\odot\!+\!1}\biggr)^2
 \!\!\frac{1}
 {1+\alpha_C^\excess\alpha_S/(8\pi^2\varepsilon_0^2R_C^6)}\,,\\
\label{betaSC2}
\beta_{S+C}^\excess=\beta_C^\excess
+\beta_S\mu_\odot\,
 \biggl(\frac{3}{2\mu_\odot\!+\!1}\biggr)^2
 \!\!\frac{1}{1+\beta_C^\excess\beta_S\mu_0^2/(8\pi^2R_C^6)}\,.
\end{gather}
As we see, the response of the sphere plus cavity system to an
electromagnetic field is due to reflection at the cavity surface from
the outside ($\alpha_C^\excess$,$\beta_C^\excess$) plus reflections at
the sphere ($\alpha_S^\excess$,$\beta_S^\excess$), where the
local-field correction factors in large round brackets account for the
transmission of the field into and out of the cavity and the
denominators account for multiple reflections between the cavity and
sphere surfaces. 

Note that in our leading-order approximation in terms of the cavity
and sphere radii, the reflective properties of the cavity and the
sphere as encoded via their dipole polarisabilities and
magnetisabilities are proportional to the third power of these radii.
On the contrary, the transmission properties of the cavity as
described by the local-field correction factors become independent of
$R_C$ within leading order of $kR_C$.


\subsection{Casimir--Polder potential}
\label{Pot}

Consider a polarisable and magnetisable ground-state atom that
interacts with a small magnetodielectric sphere in an arbitrary
environment. We assume the atom--sphere separation $r_{AS}$ to be
much greater than the effective sphere and cavity radii,
$\sqrt{\varepsilon_S\mu_S}\,R,%
\sqrt{\varepsilon_\odot\mu_\odot}\,R_C\ll r_{AS}$. The frequency
integral in Eq.~(\ref{Ue}) being typically limited to values $\xi\le
c/r_{AS}$, the assumptions $|k_SR|,|kR_C|\ll 1$ made in the previous
section~\ref{Green} hold. We can hence use our results for the
electric and magnetic Green tensors $\ten{G}_S^{(1)}$ and
$\bm{\nabla}\vprod\ten{G}^{(1)}_S\vprod\overleftarrow{\bm{\nabla}}'$ 
in the presence of a small magnetodielectric sphere to calculate the
atom--sphere potential.


\subsubsection{Full sphere}
\label{CPsphere}

Substituting Eq.~(\ref{Gsresultgeneral}) into Eq.~(\ref{Ue}), the
interaction of an electric atom with a magnetodielectric sphere is
described by the potential
\begin{equation}
\label{Ue2}
U_e(\mathbf{r}_A, \mathbf{r}_S)=U_{ee}(\mathbf{r}_A, \mathbf{r}_S)+
U_{em}(\mathbf{r}_A, \mathbf{r}_S),
\end{equation}
with
\begin{multline}
\label{Uee}
U_{ee}(\mathbf{r}_A, \mathbf{r}_S)=-\frac{\hbar\mu_0^2}{2\pi}\int _0
 ^\infty \mathrm{d}\xi\,\xi^4 \alpha_A(i\xi)
\biggl[\frac{3\varepsilon_A(i\xi)}{2\varepsilon_A(i\xi)+1}\biggr]^2\\
\times\alpha_S(i\xi)\varepsilon_\odot(i\xi)\mathrm{Tr}\bigl[
 \ten{G}(\mathbf{r}_A,\mathbf{r}_S,i\xi)\cdot
 \ten{G}(\mathbf{r}_S,\mathbf{r}_A,i\xi)\bigr]
\end{multline}
and
\begin{multline}
\label{Uem}
U_{em}(\mathbf{r}_A, \mathbf{r}_S)=-\frac{\hbar\mu_0^2}{2\pi}
 \int _0 ^\infty\mathrm{d}\xi\,\xi^2 \alpha_A(i\xi)
\biggl[\frac{3\varepsilon_A(i\xi)}{2\varepsilon_A(i\xi)+1}\biggr]^2\\
\times
\frac{\beta_S(i\xi)}{\mu_\odot(i\xi)}\mathrm{Tr}\Bigl[\ten{G}
(\mathbf{r}_A,\mathbf{r}_S,i\xi)\!\times\!\Lnabla_S
\!\cdot\!\Nabla_S\!\times\!\ten{G}(\mathbf{r}_S,\mathbf{r}_A,i\xi)\Bigr]
\end{multline}
being associated with the electric and magnetic properties of the
sphere, respectively. Similarly, combining
Eqs.~(\ref{Gsmresultgeneral}) and (\ref{Um}) gives the CP
interaction of a magnetic atom and a magnetodielectric sphere
\begin{equation}
\label{Um2}
U_m(\mathbf{r}_A,\mathbf{r}_S)=U_{me}(\mathbf{r}_A,\mathbf{r}_S)+U_{mm}
(\mathbf{r}_A,\mathbf{r}_S)
\end{equation}
with
\begin{multline}
\label{Ume}
U_{me}(\mathbf{r}_A, \mathbf{r}_S)=-\frac{\hbar\mu_0^2}{2\pi}\int_0^\infty
\!\!\mathrm{d}\xi\,\xi^2 \beta_A(i\xi)
\biggl[\frac{3}{2\mu_A(i\xi)+1}\biggr]^2\\
\times\alpha_S(i\xi)\varepsilon_\odot(i\xi)\mathrm{Tr}\Bigl\{\bigl[
\Nabla_{\!A}\!\times\!\ten{G}(\mathbf{r},\mathbf{r}_S,i\xi)\bigr]\\
\cdot\!\bigl[\ten{G}(\mathbf{r}_S,\mathbf{r}_A,i\xi)
\!\times\!\Lnabla_{\!A}\bigr]\Bigr\}
\end{multline}
and
\begin{multline}
\label{Umm}
U_{mm}(\mathbf{r}_A,\mathbf{r}_S)=-\frac{\hbar\mu_0^2}{2\pi}\int_0^\infty
\!\!\mathrm{d}\xi\beta_A(i\xi)
\biggl[\frac{3}{2\mu_A(i\xi)+1}\biggr]^2\\
\times\frac{\beta_S(i\xi)}{\mu_\odot(i\xi)}\mathrm{Tr}\Bigl\{\bigl[
\Nabla_{\!A}\!\times\!\ten{G}(\mathbf{r}_A,\mathbf{r}_S,i\xi)
\!\times\!\Lnabla_S\bigr]
\\
\cdot\!\bigl[\Nabla_S\!\times\!\ten{G}(\mathbf{r}_S,\mathbf{r}_A,i\xi)
\!\times\!\Lnabla_{\!A}\bigr]\Bigr\}.
\end{multline}
Note that the electric and magnetic properties of the sphere
completely decouple and give rise to the separate potentials $U_{ee}$,
$U_{me}$ and $U_{em}$, $U_{mm}$, respectively. However, this is only
true in the small-sphere limit considered here.

As proven in Refs.~\cite{Buhmann2008a,Buhmann2009}, the local-field
corrected total CP potential of a magnetodielectric ground-state atom
in the presence of an arbitrary arrangement of bodies as given by
Eqs.~(\ref{Ue}) and (\ref{Um}) is always duality invariant. By using
the transformation rules~(\ref{dgftrans1})--(\ref{dgftrans4}) for the
Green tensor together with $\alpha^{\star\dual}=\beta^\star/c^2$,
$\beta^{\star\dual}=c^2\alpha^\star$, one sees that duality invariance
also holds for the special case of a sphere, where
$U_{ee}(\mathbf{r}_A,\mathbf{r}_S)^\circledast$
$=U_{mm}(\mathbf{r}_A,\mathbf{r}_S)$ and
$U_{em}(\mathbf{r}_A,\mathbf{r}_S)^\circledast$
$=U_{me}(\mathbf{r}_A,\mathbf{r}_S)$. This property is ensured by the
presence of the factors $\varepsilon_\odot$ and $1/\mu_\odot$ in
Eqs.~(\ref{Uee}), (\ref{Uem}), (\ref{Ume}) and (\ref{Umm}).

It is instructive to compare our findings with the vdW interaction
between two magnetodielectric ground-state atoms $A$ and $B$ in the
presence of an arbitrary magnetodielectric environment 
\cite{Sambale2007,Safari2008}. In order to reproduce those results,
one has to perform the substitutions
\begin{align}
\label{compare}
&\alpha_S\varepsilon_\odot 
\rightarrow
\alpha_B\biggl(
\frac{3\varepsilon_B}{2\varepsilon_B+1}\biggr)^2
\\
\label{comp2}
&\frac{\beta_S}{\mu_\odot}
\rightarrow\beta_B\biggl(
\frac{3}{2\mu_B+1}\biggr)^2\,. 
\end{align}
The difference between the case of a sphere [left hand sides of
Eqs.~(\ref{compare}) and (\ref{comp2})], and an atom [right hand
sides], are due to the microscopic/macroscopic nature of the two
objects. The sphere consists of a large number of atoms whose
magnetoelectric response can be described by an average permittivity
and permeability. In this macroscopic picture, the sphere is in
immediate contact with the surrounding medium (also characterised by a
permittivity and permeability), which leads to the factors
$\varepsilon_\odot$ and $1/\mu_\odot$. In contrast, an atom is a
microscopic object. In the microscopic picture, the interspace between
the atom and the neighboring medium atoms needs to be taken into
account; it gives rise to the local-field correction factors on the
right hand sides of Eqs.~(\ref{compare}) and (\ref{comp2}).

The second difference between the two cases is in the different
explicit forms of polarisability and magnetisability. For a sphere,
they are given in terms of the permeability and permittivity of the
sphere in comparison to those of the surrounding medium,
cf.~Eqs.~(\ref{alpha}) and (\ref{beta}); they can be either positive
or negative. For an atom, polarisability and magnetisability depend on
the transition frequencies and dipole matrix elements [recall
Eqs.~(\ref{eq:alpha}) and (\ref{eq:beta})]; they are strictly positive
on the positive imaginary frequency axis.


\subsubsection{Sphere inside an Onsager cavity}
\label{CP3L}

In order to interpolate between the two extreme cases of a single atom
and a sphere consisting of a very large number of atoms, we now consider
the CP interaction of an atom with a sphere of radius $R$ that is
separated from the surrounding medium by a spherical free-space cavity
of radius $R_C$, as introduced in Sec.~\ref{sub3L}. Since
expressions of the type (\ref{Gsresultgeneral}) and
(\ref{Gsmresultgeneral}) remain valid, their substitution into
Eqs.~(\ref{Ue}) and (\ref{Um}) again leads to
Eqs.~(\ref{Ue2})--(\ref{Umm}), where now $\alpha_{S+C}^\excess$ and
$\beta_{S+C}^\excess$ as given by Eqs.~(\ref{alphaSC}) and
(\ref{betaSC}) appear in place of $\alpha_S^\excess$ and
$\beta_S^\excess$. 

In our model, the sphere may consist of an arbitrary number of atoms,
while the cavity implements the interspace between the sphere's atoms
and the surrounding medium atoms. As seen from Eqs.~(\ref{alphaSC2}) 
and (\ref{betaSC2}) for the polarisability and magnetisability of the
sphere plus cavity system, the sphere is represented by its
free-space polarisability and magnetisability, while the interspace
gives rise to the cavity excess polarisability and magnetisability.
In the purely electric case, the sphere gives rise to attractive
forces while the cavity leads to a reduction of these forces. 

For a sphere that consists of a very large number of atoms, the
interspace between the sphere and medium atoms becomes irrelevant. In
this case, which is implemented by the limit $R\to R_C$, the system's
polarisability and magnetisability become equal to the excess
polarisability and magnetisability of a full sphere (recall
Sec.~\ref{sub3L}), for which we recover Eqs.~(\ref{Ue2})--(\ref{Umm})
in their original form. 

In the opposite extreme case of a sphere that consists of only
very few atoms, the interspace becomes very large in comparison to the
sphere, $R\ll R_C$. In this limit, the effect of multiple scattering
between the surfaces of sphere and cavity becomes negligible, and the
polarisability~(\ref{alphaSC2}) and magnetisability~(\ref{betaSC2})
reduce to
\begin{gather}
\label{alphaSCsmall}
\alpha_{S+C}^\excess=\alpha_C^\excess
+\frac{\alpha_S}{\varepsilon_\odot}\,
 \biggl(\frac{3\varepsilon_\odot}{2\varepsilon_\odot\!+\!1}\biggr)^2
 \,,\\
\label{betaSCsmall}
\beta_{S+C}^\excess=\beta_C^\excess
+\beta_S\mu_\odot\,
 \biggl(\frac{3}{2\mu_\odot\!+\!1}\biggr)^2\,.
\end{gather}
When the sphere consists of only a single atom $B$, the
Clausius--Mossotti laws \cite{0001}
\begin{equation}
\label{Clausius1}
\frac{\varepsilon_S-1}{\varepsilon_S+2}
=\frac{\alpha_B}{3\varepsilon_0V}\,,\qquad
\frac{\mu_S-1}{\mu_S+2}
=\frac{\mu_0\beta_B}{3V}
\end{equation}
[$V=(4\pi/3)R^3$], together with Eqs.~(\ref{alphafree}) and
(\ref{betafree}), show that $\alpha_S=\alpha_B$ and 
$\beta_S=\beta_B$. When neglecting the backscattering from the
outside surface of the cavity, we obtain
\begin{align}
\label{reduce}
&\alpha_{S+C}\varepsilon_\odot 
=\alpha_B\biggl(
\frac{3\varepsilon_B}{2\varepsilon_B+1}\biggr)^2\,,
\\
\label{reduce2}
&\frac{\beta_{S+C}}{\mu_\odot}
=\beta_B\biggl(
\frac{3}{2\mu_B+1}\biggr)^2\,. 
\end{align}
($\varepsilon_\odot=\varepsilon_B$, $\mu_\odot=\mu_B$) and
substitution into Eqs.~(\ref{Ue2})--(\ref{Umm}) leads to the
local-field corrected two-atom potentials
\cite{Sambale2007,Safari2008}. It is in this limit
$R\ll R_C$, $\sqrt{\varepsilon_\odot\mu_\odot}\,R_C\ll r_{AS}$ that
the potential depends on the cavity radius only via its transmission
properties and therefore becomes independent of $R_C$, recall the
dicussion below Eq.~(\ref{betaSC2}).

For intermediate radii $R$, our sphere plus cavity model gives a good
description of the interaction of a single atom with molecules or
intermediate-size spherical clusters of atoms. Note that in this case
the potential explicitly depends on both the sphere and cavity radii.
The theory could be applied, e.g., in cell biology, to study the
vdW-force induced transfer of an atom or a very small molecule from
one cell into another where it is attracted to another bigger
(spherical) cell component or molecule. Note that local-field effects
are automatically included.

\begin{figure}[!t!]
\begin{center}
\includegraphics[width=\linewidth]{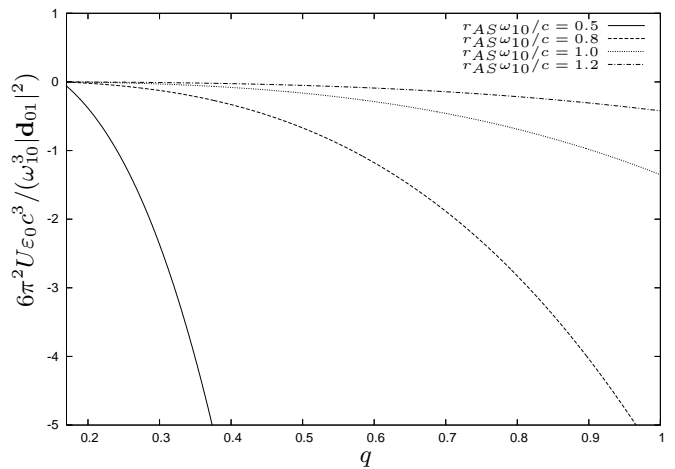}
\end{center}
\caption{CP potential $U(r_{AS})$ of a nonmagnetic atom in front of
a dielectric sphere in an empty cavity embedded in bulk material vs.
$q=R/R_\mathrm{C}$ for different atom--sphere separations
$r_{AS}\omega_{10}/c$. Other parameters are
$\omega_{T}/\omega_{10}=1.03$,
$\omega_{TS}/\omega_{10}=1.0$, 
$\omega_{PS}/\omega_{10}=6.0$,
$\omega_{P}/\omega_{10}=0.1$, 
$\gamma_{(S)}/\omega_{10}=0.001$.
} 
\label{figUeed}
\end{figure}

\begin{figure}[!t!]
\begin{center}
\includegraphics[width=\linewidth]{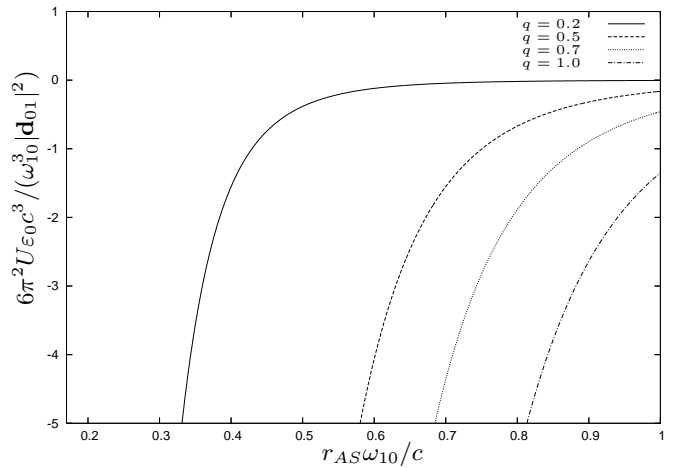}
\end{center}
\caption{$U(r_{AS})$ vs. $r_{AS}\omega_{10}/c$ for different
ratios $q$. Other parameters are the same as in Fig.~\ref{figUeed}.
} 
\label{figUeeR}
\end{figure}

As an example, let us consider the CP interaction of a nonmagnetic
atom with a purely electric sphere ($U_{ee}$) in a bulk medium
($\varepsilon_\odot=\varepsilon_A=\varepsilon$). Substituting the
required Green tensor~(\ref{bulk}) into Eq.~(\ref{Uee}), one easily
finds
\begin{multline}
\label{Uee3}
U(\mathbf{r}_A,\mathbf{r}_S)
=-\frac{\hbar}{16\pi^3\varepsilon_0^2r_{AS}^6}
 \int_0^\infty \mathrm{d}\xi\biggl[\frac{3\varepsilon(\mi\xi)}
{2\varepsilon(\mi\xi)+1}\biggr]^2\\
\times\alpha_A(\mi\xi)\varepsilon(\mi\xi)\alpha_{S+C}^\excess(\mi\xi)
g\bigl[\sqrt{\varepsilon(\mi\xi)}\,\xi r_{AS}/c\bigr]
\end{multline}
($r_{AS}=|\mathbf{r}_A-\mathbf{r}_S|$) with
\begin{equation}
\label{g}
g(x)=e^{-2x}(3+6x+5x^2+2x^3+x^4).
\end{equation}
Figure \ref{figUeed} shows this potential for a two-level atom as a
function of the relative sphere radius $q=R/R_C$ for various atom--sphere
separations. We have used single-resonance models for the
permittivities of the sphere and the medium,
\begin{equation}
\label{Drude1}
\varepsilon_{(S)}(\omega)=1+\frac{\omega_{P(S)}^2}
{\omega_{T(S)}^2-\omega^2-\mi\omega\gamma_{(S)}}\;.
\end{equation}
In Fig.~\ref{figUeeR}, we show the potential $U_{ee}$ as a
function of the atom-sphere separation for different relative sphere
radii $q=R/R_C$. Both figures show that for the constant
$\varepsilon_S$ considered here, larger spheres with their
corresponding larger polarisabilities lead to stronger vdW attraction 
between the atom and the sphere.
%
%


\section{Summary and conclusions}
\label{Sum}

We have studied the CP interaction of an atom with a small
magnetodielectric sphere in an arbitrary magnetoelectric environment.
Employing a point-scattering technique, we were able to express the
Green tensor in the presence of the sphere as a simple function of the
Green tensor of the environment. Using this result, we have found
closed general expressions for the CP potential of a magnetoelectric
atom interacting with a small magnetodielectric sphere which depend
on the sphere's polarisability and magnetisability. A comparison with
the vdW potential between two ground-state atoms in the presence of
the background medium has revealed how the different
macroscopic/microscopic natures of atom versus sphere manifest
themselves in the dispersion potentials: The immediate contact of a
macroscopic sphere with the surrounding medium leads to the appearance
of the permittivity and inverse permeability of the medium, whereas
the coupling of the local electromagnetic field to the microscopic
atom gives rise to local--field correction factors. 

In order to interpolate between these two limiting cases, we have
studied the potential of an atom with a sphere of variable radius
located inside an Onsager cavity. The cavity represents the
interspace between the particles contained in the sphere and those
constituting the surrounding medium, so that our model can be used to
study molecular systems of arbitrary size. Using similar techniques,
we have derived the potential of an atom interacting with the 
sphere plus cavity system. We have shown that our result reduces to
the atom--sphere or atom--atom potentials in the two limiting cases of
the sphere radius being much smaller than or equal to the cavity
radius. As an example we have considered the CP interaction between an
electric atom and such a sphere plus cavity system, finding that the
attractive potential diminishes as the sphere becomes smaller at
fixed cavity radius.

Our point-scattering method can also be used to calculate the Casimir
force on a small sphere in an arbitrary environment and, in
particular, the Casimir force between two small spheres. This problem
will be subject of future work. A similar approach could further be
applied to dispersion interactions involving thin cylinders.


\acknowledgments

This work was supported by the UK Engineering and Physical Sciences Research 
Council and the ESF Research Network CASIMIR. S.Y.B. is grateful for
support by the Alexander von Humboldt Foundation.

\end{document}